\documentclass{article}

\usepackage{times}
\usepackage[xdvi,dvips]{epsfig}

\usepackage{supertabular}

\begin{document}

\pagestyle{empty}

{
\flushright{{\LARGE \bf Bethe Logarithms for Rydberg States:}\\[0.5ex]
{\LARGE \bf Numerical Values for $n \leq 200$}}

\vspace{.5in}

\large
\flushright{Ulrich D.~Jentschura \\
{\em Max--Planck--Institut f\"{u}r Kernphysik} \\
{\em Heidelberg, Germany and} \\
{\em NIST Division 842: Atomic Physics,} \\
{\em Physics Laboratory, USA}}

\flushright{Peter J. Mohr\\
{\em NIST Division 842: Atomic Physics,} \\
{\em Physics Laboratory, USA} }
\vfill
}

\epsfig{file=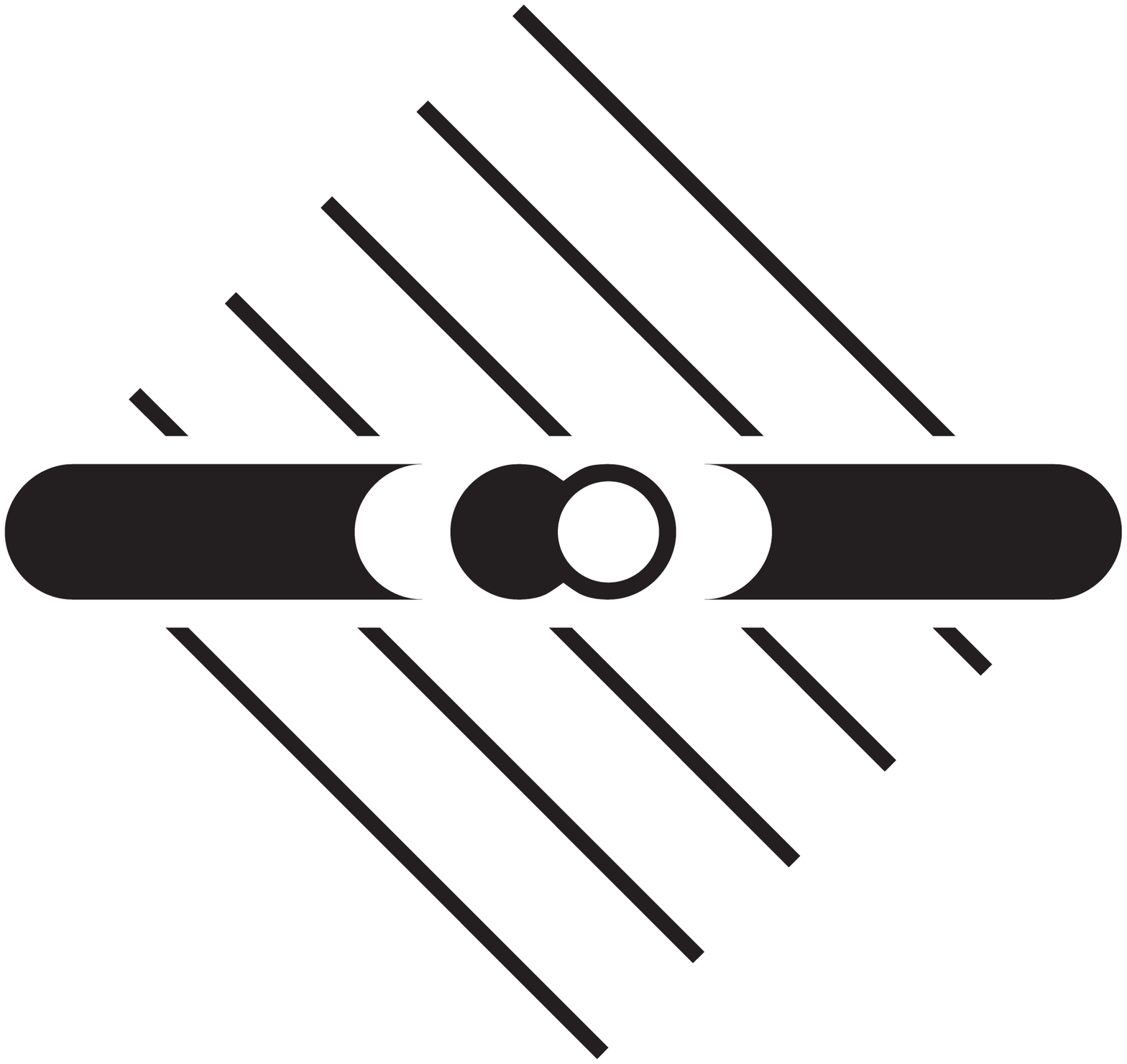,height=5.0cm}
\hfill{}
\epsfig{file=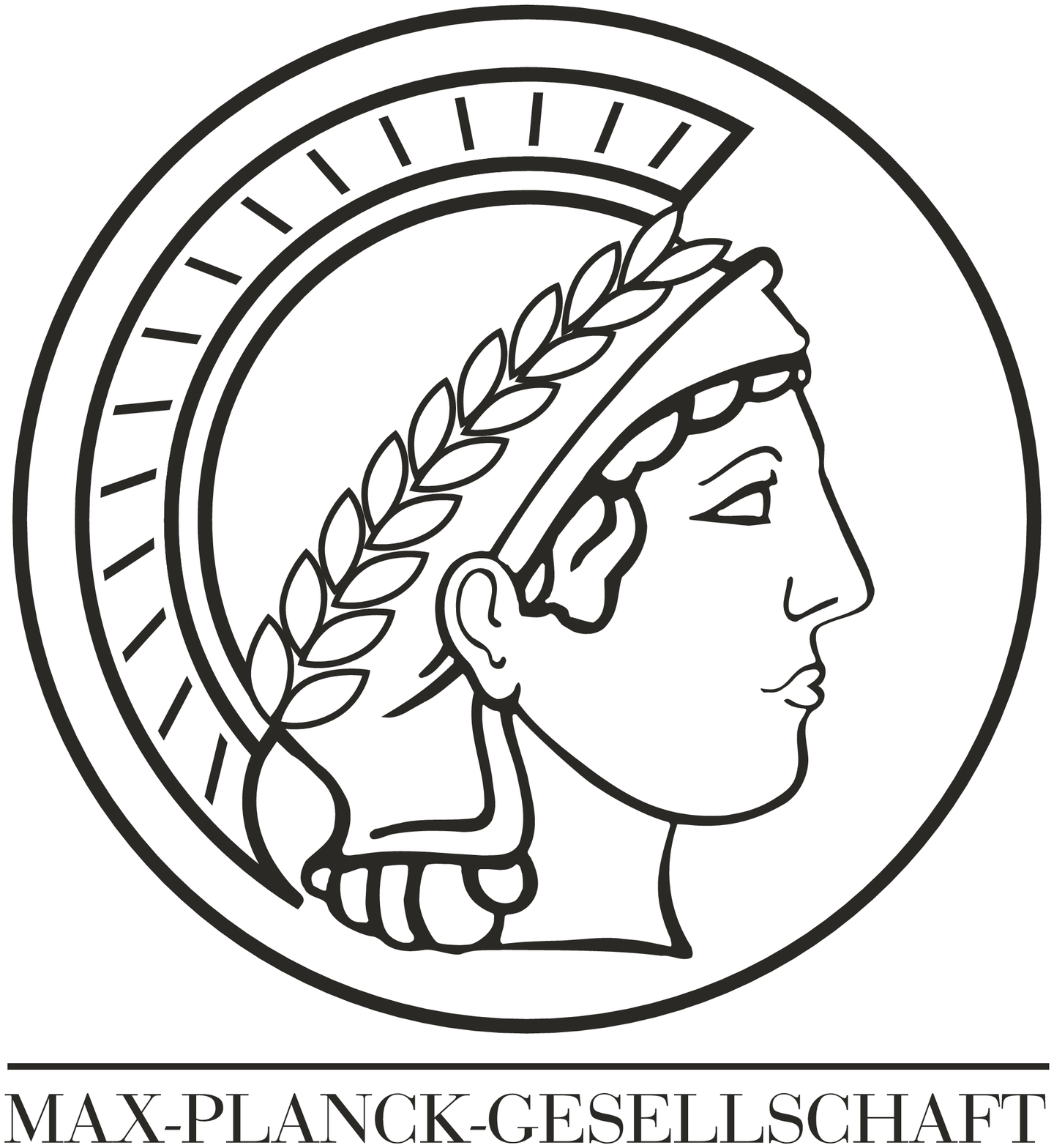,height=5.0cm}
\epsfig{file=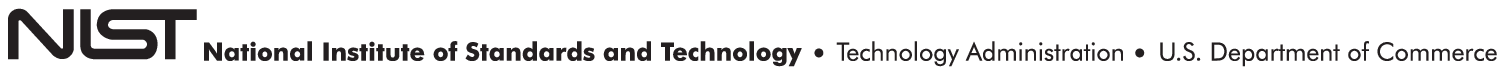,width=\textwidth}

\newpage

\pagestyle{plain}

\twocolumn

\noindent
{\it This document provides reference data for all
Bethe logarithms of hydrogenic
bound states with principal quantum numbers $n \leq 200$.
The calculation is carried out for nonrelativistic 
hydrogenic Rydberg-state wave functions.
The values are provided for each angular momentum $l$,
in increasing order of the principal quantum number $n$,
starting from $n=l+1$.

A single table is used, which is continued on a number
of pages. All the 20100 
hydrogenic bound states with principal quantum 
numbers $n \leq 200$ are treated.

The numerical data are provided with an accuracy
of nine decimals. All decimal figures shown in the table are
significant.

The size of the following table is somewhat 
lengthy. In view of the lack of 
analytic expressions for the Bethe logarithm of excited 
hydrogenic states, there is currently no replacement
for an explicit listing of all relevant values,
if one would like to reliably understand radiative corrections
for hydrogenic Rydberg levels with principal quantum numbers 
$n \leq 200$.

One of the most comprehensive compilation of Bethe logarithm 
recorded so far in the literature has covered all 
principal quantum numbers up to
$n \leq 20$ [see G. W. F. Drake and R. A. Swainson, 
Phys. Rev. A {\bf 41}, 1243 (1990)]. 
The motivation for extending the previously 
available calculations to higher principal quantum 
numbers is given by recent experiments and described 
in the accompanying e-print quant-ph/0504001.}

\begin{center}
\tablefirsthead{%
\hline\hline 
\multicolumn{1}{c}{$n$} &
\multicolumn{1}{c}{$l$} &
\multicolumn{1}{c}{$\ln k_0(n,l)$}\\
\hline}
\tablehead{%
\hline\hline
\multicolumn{3}{c}{continued from previous page/column} \\
\hline
\multicolumn{1}{c}{$n$} &
\multicolumn{1}{c}{$l$} &
\multicolumn{1}{c}{$\ln k_0(n,l)$}\\
\hline}
\tabletail{%
\hline
\multicolumn{3}{c}{continued on next column/page} \\
\hline
\hline}
\tablelasttail{\hline\hline}
\begin{supertabular}{ccc}
\input{tex.full.num}
\end{supertabular}
\end{center}

\end{document}